\documentclass[10pt,a4paper,nofootinbib]{article}
\pdfoutput=1
\usepackage{jcappub}
\usepackage{graphicx}
\usepackage{longtable,comment}
\usepackage{amsmath}
\usepackage{amssymb}
\usepackage{hyperref}
\usepackage{dcolumn}
\usepackage{multirow}
\usepackage{cancel}
\usepackage[rightcaption]{sidecap}
\usepackage[normalem]{ulem}
\usepackage{cancel}
\usepackage{xcolor}
\usepackage{sidecap}
\usepackage{caption}
\usepackage{subcaption}

\newcommand{\refeq}[1]{Eq.~(\ref{eq:#1})}
\newcommand{\tn}{\textnormal}

\newcommand{\pd}{\partial}
\newcommand{\mpl}{M_\tn{pl}}
\newcommand{\mn}{{\mu\nu}}

\newcommand{\vac}{\tn{vac}}
\newcommand{\vev}{\tn{vac}}
\newcommand{\go}{|\gamma-1|}
\newcommand{\rss}{r_\tn{SS}}
\newcommand{\qs}{\tn{QS}}

\def\DGO{\Delta_{|\gamma-1|}}

\begin{document}
\title{Tsunamis and Ripples: Effects of Scalar Waves on Screening in the Milky Way}

\author[a]{Hiu Yan Sam Ip,} 

\author[a]{Fabian Schmidt}

 \affiliation[a]{Max-Planck-Institut f\"ur Astrophysik, Karl-Schwarzschild-Str. 1, 85741 Garching, Germany}

 \emailAdd{iphys@mpa-garching.mpg.de}
 \emailAdd{fabians@mpa-garching.mpg.de}

 \abstract{
   Modified gravity models which include an additional propagating degree of freedom are typically studied in the quasi-static limit, where the propagation is neglected, and the wave equation of the field is replaced with a Poisson-type equation. Recently, it has been proposed that, in the context of models with symmetron- or chameleon-type screening, scalar waves from astrophysical or cosmological events could have a significant effect on the screening of the Solar System, and hence invalidate these models. Here, we quantitatively investigate the impact of scalar waves by solving the full field equation linearised in the wave amplitude. In the symmetron case, we find that the quantitative effect of waves is generally negligible, even for the largest amplitudes of waves that are physically expected. In order to spoil the screening in the Solar System, a significant amount of wave energy would have to be focused on the Solar System by arranging the sources in a spherical shell centred on Earth. In the chameleon case, we are able to rule out any significant effects of propagating waves on Solar System tests. 
 }

\maketitle

\section{Introduction}
\label{sec:intro}
In scalar-tensor theories of gravity, screening mechanisms are necessary to hide the extra degrees of freedom locally. This allows consistency with the Solar System tests of gravity that agree with General Relativity to high precision. We will focus on the Post-Newtonian Parameter $|\gamma-1|$ here, which measures light bending by the Sun due to  a change in how much each unit of mass warps spacetime, constrained by radio Doppler data from Cassini to be $< 2\times 10^{-5}$ \cite{cassini}. Further, we focus on screening of the chameleon \cite{Khoury03} and symmetron \cite{khourysym} types here.  
Until recently, these screening mechanisms have been studied under the quasi-static approximation, where the Klein-Gordon equation [which includes a potential $V(\phi)$] reduces to an elliptic Poisson-type equation, prohibiting any wave propagation.

Llinaries and Mota \cite{Llinares:2013jua} went beyond the quasi-static approximation and found scalar waves in non-quasi-static cosmological simulations for the  symmetron model. These appear because, at a certain cosmological epoch, the field enters a symmetry-breaking phase. This leads to large domains where the field is in different vacua, which can then rapidly transition as they come into contact. Together with Hagala \cite{mota}, they further numerically studied how a \emph{spherical incoming wave} of the  symmetron field centred on the Milky Way halo affects the screening of the Solar System, which is dominated by the halo's potential well. Astrophysical and cosmological sources for such waves were suggested.
They concluded that the amplitude of the fifth force and $|\gamma-1|$ can consequently increase by several orders of magnitude, potentially reaching $>10^{-5}$. In other words, they concluded that screening can be significantly disrupted by scalar radiation, causing  previously viable  models to violate current bounds.
Such effects could in principle be relevant for any modified gravity theories with extra degrees of freedom with wave-type equations of motion. Potential sources of this scalar radiation include the Mpc-scale waves generated in the process of structure formation in the case of the symmetron model, as found by \cite{Llinares:2013jua}. Specifically, they found a peak amplitude of waves (at $a\approx 0.4$, where $a$ is the cosmological scale factor) corresponding to a fractional field perturbation of $\delta\phi/\phi \sim 0.03$, although typical wave amplitudes are significantly smaller. 
Further, in both symmetron and chameleon models, any unscreened astrophysical object that collapses to form a screened object has to radiate away its scalar charge. This in particular includes the collapse of massive stars to neutron stars or black holes (which we will hereafter refer to as Supernovae). In this case, assuming that the scalar field couples with approximately gravitational strength, the wave amplitude can be estimated to be of order $\phi \sim \Phi_\star R_\star/r$, where $\Phi_\star$ is the surface gravitational potential of the star and $R_\star$ is the stellar radius (both before collapse), and $r$ is the distance to the source. Note that only sources in unscreened regions are relevant, since otherwise the star does not carry scalar charge even before collapse. This means that the sources considered here are outside of the screened part of the Milky Way halo, so that $r \gtrsim 100$~kpc.

In this paper, we shall obtain analytical solutions for the influence of spherical as well as planar incoming waves, in order to obtain better physical insight into this scenario. We linearise the system and consider a tophat halo in order to obtain closed-form solutions. While linear theory breaks down for incoming waves of very large amplitudes, we can nonetheless gain physical insight. In particular, purely geometrical effects are expected to qualitatively hold for larger wave amplitudes as well. 
We expect planar waves to be generally the most physically relevant wave configuration. Clearly, given the large distance compared to the source size, the planar-wave assumption should be very accurate for astrophysical events. For symmetron waves of cosmological origin, this is less clear, as they are produced throughout the Universe and we are not necessarily in the far-field limit. Nevertheless, it seems reasonable to expect that the cosmological symmetron radiation field can be represented as a superposition of plane waves with random wavevectors and phases. We will mostly consider the spherical case in order to make the connection to the setup considered in \cite{mota}. This case corresponds to a spherical shell of emitters centred on the Milky Way halo with a \emph{single coherent phase}.

\begin{table}[ht]
\caption{Physical variables used throughout the paper. The lower part of the table lists the dark matter halo parameters adopted throughout.}\label{tab:gen}
\begin{tabular}{ p{2cm} p{9cm} p{3cm}   }
 \hline
Symbol & Definition & Value\\
 \hline
\multicolumn{3}{|c|} {\textbf{Global quantities:}}\\\hline
$g_\mn$&  Einstein frame (geometric) metric &\\ 
 $\tilde{g}_\mn$ & Jordan frame (physical) metric, $\tilde{g}_\mn= C^2(\phi)  g_\mn$&\\
$R$ &  Ricci scalar &\\
 $\mathcal{L}_m$& Lagrangian for  matter fields $\psi$& \\
 $h$ & Dimensionless Hubble parameter (today), $\frac{H_0}{100 \tn{ km s}^-1}$& $0.67$ \cite{h}\\
 $M_{pl}$ & Reduced Planck mass, $\sqrt{\hbar/8\pi G}$   & $2\times 10^{18} \,\tn{ GeV}/c^2 $\\
 $\rho(x)$ & Total matter density& \\
$\rho_c$ & Critical density of the universe& $9.5\times10^{-27}\, \tn{kg}\,\tn{m}^{-3}$\\
$\rho_m:= \Omega_m \rho_c$  & Background matter density of the universe & $2.6\times 10^{-27}\, \tn{kg}\,\tn{m}^{-3}$\\
$a$ &  Expansion factor of the universe & $1$ today\\
$H$&Hubble parameter&\\
$\gamma$ &PPN parameter measuring the space-curvature produced by a unit rest mass \cite{will}, tracked by the Cassini probe. & $|\gamma-1|<2\times 10^{-5}$  \cite{cassini}\\
$\DGO$ & Fractional deviation of the PPN parameter from the quasi-static case: $ \left||\gamma-1|_\tn{non-QS}-|\gamma-1|_\qs\right| / |\gamma-1|_\qs $&\\
$P_L$& Legendre polynomial of order $L$&\\
$j_L$ &Spherical Bessel function of order $L$&\\
  \hline\multicolumn{3}{|c|} {\textbf{Halo parameters:}}\\\hline
$R_h$ &Milky Way halo's virial radius taken to be the tophat halo radius & $200 \,\tn{kpc}$\\
$r_\tn{SS}$& Radial distance of the Solar System from the center of the Milky Way halo &$8 \,\tn{kpc}$\\
$\rho_h=200 \rho_c$ & Average interior density of the Milky Way halo, obtained from virial mass and radius & $1.9\times 10^{-24}\, \tn{kg}\,\tn{m}^{-3}$ \\
$M_h$ & Mass of the MW halo & $8.4\times 10^{11}M_\odot$ \\
$\Phi_\tn{in}$ & Gravitational potential inside the halo&  $\frac{G M_h (3 R_h^2-r^2)}{2 R_h^3}$ 
  \\
  \hline\\
\end{tabular}
\end{table}

Unfortunately, we will generally find a very small impact of scalar waves on the Solar System, even when considering wave amplitudes at the upper limit of what is physically expected. 

This paper is organized as follows:
Section \ref{sec:sys} offers a mathematical formulation of the system in the symmetron model. Section \ref{sec:sol} provides and analyses the solutions for the different types of incoming waves. Section \ref{sec:cham} investigates the same scenario in the chameleon case.   Section  \ref{sec:conc} concludes the paper and offers directions for  future research.
Table \ref{tab:gen} lists the notation and quantities used throughout the paper.
Table \ref{tab:sym} lists the quantities, including their definitions, notations and adopted values, for the symmetron model, while Table \ref{tab:cham} lists those for the chameleon model.

\section{The symmetron case}
\label{sec:sys}

We shall consider a tophat Milky Way halo subjected to spherical and planar incoming waves.  We do so by linearising the Klein-Gordon equation, including potential, in the time-dependent perturbation $\delta \phi$. This yields a linear Klein-Gordon equation with an effective mass that depends on radius. In the case of the tophat halo, the mass is given by a step function, assuming its cosmological value outside of the halo, and a larger value inside. 

The symmetron model is a special case of the general scalar-tensor action for canonical scalar fields,
\begin{equation}\label{eq:STA}
S=\int d^4 x \sqrt{-g} \left[\frac{\mpl^2}{2}R -\frac{1}{2}g^\mn \pd_\mu \phi \pd_\nu \phi -V(\phi)\right] +\int d^4 x  \sqrt{-\tilde{g}}  \mathcal{L}_m (\psi,\tilde{g}_\mn)\,,
\end{equation}
where $\psi$ denotes the matter fields and $\tilde g_\mn$ is the Jordan-frame metric. In the symmetron case, the latter is given by
\begin{equation}
\tilde{g}_\mn= C^2(\phi) g_\mn,\quad C(\phi) =1+ \frac{\phi^2}{2 M^2}+ \mathcal{O}\left(\frac{\phi^4}{M^4}\right)  
\end{equation}
and the quartic Symmetron potential is 
\begin{equation}
V(\phi)= -\frac{1}{2}\mu^2 \phi^2 + \frac{1}{4}\lambda \phi^4 + V_0,
\end{equation}
with $\mu,\lambda, V_0$ being free parameters. Our adopted values are defined in Table \ref{tab:sym}. 

\begin{table}[b]
\caption{Quantities, including their definitions, notations and adopted values, which characterize the symmetron model adopted here.}\label{tab:sym}
\begin{tabular}{ p{2cm} p{9cm} p{3cm}   }
 \hline
Symbol & Definition & Value\\
 \hline
$\lambda_\vac$ & Symmetron range in vacuum &$0.37\, \tn{Mpc}$ \cite{mota}\\
$\mu=\frac{1}{\sqrt{2}\lambda_\vac}$ & Mass scale defined via $V(\phi)$& $1.9\,\tn{Mpc}^{-1}$\\
 $M$ & Mass scale defined in $ C(\phi) =1+ \frac{\phi^2}{ 2 M^2}+ \mathcal{O}\left(\frac{\phi^4}{M^4}\right) $& $8.2\times 10^{14} \,\tn{GeV}$\\
 $\phi_\vev = \frac{\mu }{\sqrt{\lambda}}$ & Vacuum expectation value of $\phi$ with the symmetron coupling constant set to $1 ,  \, \approx M^2 / \mpl$ & $2.8 \times 10^{11} \,\tn{GeV}$\\
$\beta=\frac{\phi_\vev M_{pl}}{M^2}$ &Dimensionless symmetron coupling constant & $1$ \cite{mota} \\ 
$a_\tn{SSB}$ & Expansion factor at the time of spontaneous symmetry breaking, such that $\rho_m |_{a=1}\,{a_\tn{SSB}}^{-3}=M^2 \mu^2$\cite{assb} & $0.5$\cite{mota}\\
 $\phi_\qs(r)$ & Background scalar field under quasi-static approximation & \\
 $\delta\phi(\vec{x},t)$ & Time-dependent field perturbation, having dropped the quasi-static approximation & \\
 $m_\tn{eff,in}$ & Effective mass of the field inside the halo & $16\tn{ Mpc}^{-1}$ \\
 $r_\tn{BC}$& Edge of the simulation in \cite{mota} &$  6.0 \, \tn{Mpc}$\\
 $A$  & Dimensionless wave amplitude  defined  at $r_{\rm BC}$& $0.01$ \cite{mota}\\
$\mathcal{A}:=A \phi_\vev $ & Wave amplitude & $2.8 \times 10^{9} \,\tn{GeV}$\\
$w$ & Frequency of incoming wave& $40\, \tn{Myr}^{-1}$\\
$k=\frac{w}{v_p}$ & Wavenumber of incoming wave& $1.3\times 10^2 \tn{ Mpc}^{-1}$\\
$v_p$ & Phase velocity of incoming wave& $\sqrt{1+\frac{m_\tn{eff}^2}{k^2}}$ \cite{1512.00615}\\
  \hline\\
\end{tabular}
\end{table}

The Einstein-frame metric, for negligible backreaction of $\phi$ on the metric which applies in the weak gravity case, is 
\begin{equation}
 ds^2 = -(1+2 \Psi) dt^2 +(1-2\Psi) a^2(t) d\vec{x}^2.
\end{equation}
At this order, $\Psi=\tilde{\Psi}$, where $\tilde{\Psi}$ is the Jordan-frame gravitational
potential.  The  PPN parameter $\gamma$ which we are interested in is, essentially, the ratio of the space-like and time-like perturbations of the Jordan-frame metric in the Solar System.  Ref.~\cite{mota} defined this as the ratio of the space-like and time-like perturbations of the Jordan-frame metric:\footnote{Here we have corrected a sign error in \cite{mota}.}
\begin{equation}
 \gamma-1 \Big|_\text{halo} = -\frac{\phi^2}{M^2} \frac{2}{\frac{\phi^2}{M^2} +2 \Psi \left(1+ \frac{\phi^2}{M^2} \right)}\,.
  \label{eq:gammahalo}
\end{equation}
Thus, this is the PPN parameter valid for the smooth spherically symmetric density profile of the halo. On the other hand, the arguably more relevant definition is the ratio of space-like and time-like perturbations generated in the metric by a test mass embedded in the halo, i.e. the Sun. Moreover, this second definition is also more consistent with the PPN framework, which is formulated around a spherically symmetric asymptotically flat spacetime, corresponding to an isolated central mass. In this case, one obtains 
\cite{qssym} [Eq.~(32) with Eq.~(18) there]
\begin{equation}
\gamma-1\Big|_\text{SS} = - \frac{4 \beta^2 (\phi/\phi_\vac)^2}{1+8 \beta^2 (\phi/\phi_\vac)^2} = - 4 \beta^2 \left(\frac{\phi}{\phi_\vac}\right)^2+\mathcal{O}\left[\left(\frac{\phi}{M}\right)^3\right]\,.
  \label{eq:gammaSS}
\end{equation}
In order to compare with the analysis of \cite{mota}, we will present results for both definition of $\gamma$ in this paper. Of course, the underlying field solution is the same. Moreover, our main conclusions hold for both choices of $\gamma$. 
We will find that the fractional difference in $\DGO$ is small ($\lesssim 0.1$) inside the halo, with the  major deviations located towards the edge of the halo, far from $\rss$.

The equation of motion of the symmetron field $\phi$,  in the absence of the quasi-static assumption, is the Klein-Gordon equation,
\begin{equation}
\Box \phi 
\approx \ddot{\phi} -\nabla^2 \phi 
=-{V}_{\tn{eff},\phi} , \quad \tn{where }
{V}_{\tn{eff}}(\phi)= V(\phi) + [C(\phi)-1]\, \rho,
\end{equation}
and $\Box$ is the D'Alembertian, a dot denotes a partial derivative with respect to cosmic time $t$, and $\rho$ is the matter density; more precisely, the field is sourced by the trace of the stress-energy tensor, but we assume pressureless matter throughout. In the second equality we have again assumed the weak-gravity and subhorizon regime.
Here and throughout, we assume the subhorizon regime today, so that we can neglect $H$ relative to $\partial_t$ and $\partial_x$, and can set $a=1$ for the scale factor. We now make the following ansatz for the field:
\begin{equation}
  \phi(\vec{x},t) = \phi_\qs(r) + \delta\phi(\vec{x},t)\,,
\end{equation}
where $\phi_\qs(r)$ is the spherically symmetric static background, and $\delta\phi$ is the wave perturbation. At zeroth and linear order in $\delta\phi$, this yields, respectively, 
\begin{align}
  \nabla^2 \phi_\qs =\:& {V}_{\tn{eff},\phi}(\phi_\qs) \\
  \label{linearsymmetron}
\pd_t^2 \delta\phi -\nabla^2 \delta\phi =\:& -m_\tn{eff,in}^2(\phi_\qs)\delta\phi , \quad \tn{where } \\
m_\tn{eff,in}^2(\phi_\qs)=\:& V_{\tn{eff},\phi\phi}(\phi_\qs) = \frac{1}{2\lambda_\vac^2}\left(\frac{3 \phi_\qs^2 }{\phi_\vev^2}-1\right)+\frac{ \rho }{M^2} \,.
\end{align}
We shall take $m_\tn{eff,out}(\phi_\qs)\approx 0$ outside the halo, assuming that the sources are within the Compton length of the field in the background. This is conservative, since including the background mass would further dampen in the waves. Inside the halo, $\left(\frac{\phi_\qs}{\phi_\vev}\right)^2\ll 1$, such that $m_\tn{eff,in}^2\approx \frac{1}{2\lambda_\vac^2}\left(\frac{a_\tn{SSB}^3 \rho_h}{\rho_m}-1\right) $, where we have introduced the expansion factor at the time of spontaneous symmetry breaking, $a_\tn{SSB}$.
This yields $m_\tn{eff,in} \approx 16 \tn{Mpc}^{-1}$. 

The incoming waves enter as boundary conditions.
A planar incoming wave can be expressed as a sum of spherical waves by the plane wave expansion:
\begin{equation}
\delta\phi (r_\tn{BC}, t)
=\tn{Re}\left[\mathcal{A}\,\tn{e}^{i(\vec{k}\cdot \vec{r}_\tn{BC}-wt)} \right]
= \tn{Re}\left[ \mathcal{A} \, e^{-i w t}\sum^\infty_{L=0} (2L+1) i^L j_L(k\, r_\tn{BC} ) P_L(\cos\theta)\right].
\end{equation}
 We choose the wavevector to be along the $z$-axis, $\vec{k} = (0,0,k)$.
A spherical incoming wave is the special case of angular independence, such that $L=0$,
\begin{equation}
\delta\phi (r_\tn{BC}, t) =  \mathcal{A}  \sin (wt).
\end{equation}

\section{Wave-induced perturbations inside the halo}
\label{sec:sol}

\begin{SCfigure}[3][t!]
\caption{Refraction of plane wave entering the halo. Note the direction of distortion due to $v_p>c$ inside the halo.}
\includegraphics[width=0.3\textwidth]{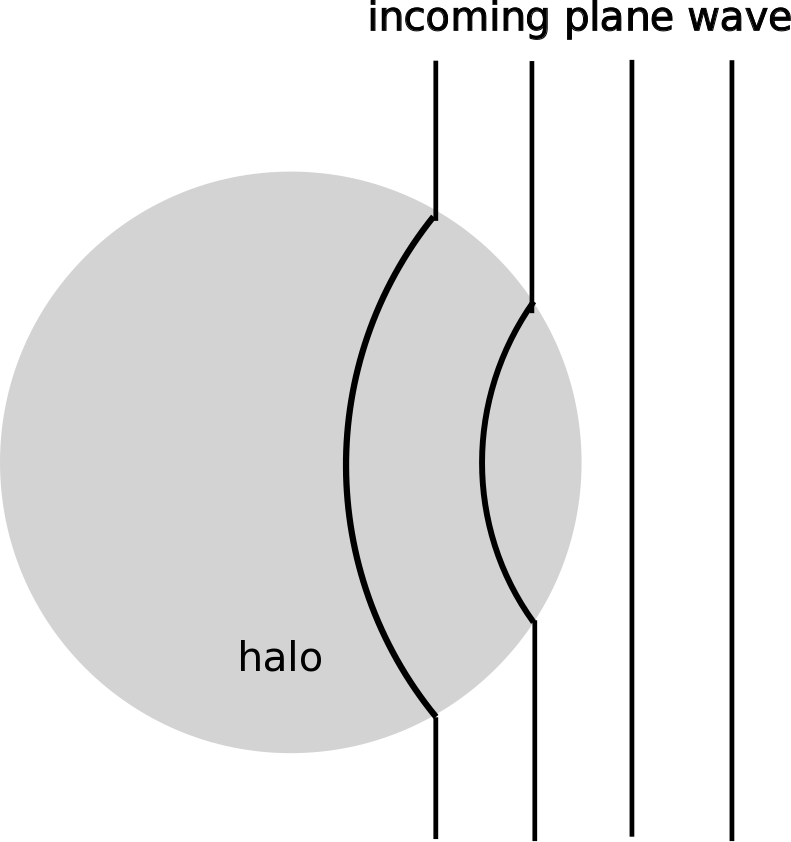}\label{fig:refraction}
\end{SCfigure}

Under the approximations explained above, the system can now be solved straightforwardly by continuously matching the spherical-wave expansions of the Klein-Gordon solutions for $m_\tn{eff,out}=0$ and $m_\tn{eff,in}$, respectively. For the incoming plane wave, we obtain
\begin{equation}\label{eq:symplanelin}
\delta \phi 
= \tn{Re}\left[\mathcal{A}   e^{-i w t}\sum_{L=0}^\infty    (2L+1) i^L \frac{ j_L(k R_h )}{ j_L(\sqrt{k^2 -m_\tn{eff,in}^2} R_h)}   \,  j_L(\sqrt{k^2-m_\tn{eff,in}^2 }\, r) P_L(\cos \theta)\right] \,.
\end{equation}
This result includes all linear optics effects such as refraction (diffraction is not relevant, as there are no barriers involved). Indeed, one expects refraction effects as the wave enters the halo due to the change in its phase velocity, $v_p=c\,\sqrt{1+\frac{m_\tn{eff,in}^2}{k^2}}>c$ \cite{1512.00615}. This means that the ratio of refractive indices $\frac{n_\tn{out}}{n_\tn{in}}>1$, resulting in convex wave fronts w.r.t. the halo center, as illustrated in Fig.~\ref{fig:refraction}. That is, the plane wave fronts are deformed to deviate even further from a spherical incoming wave. Our treatment does not include gravitational lensing. However, lensing effects are extremely small for sources at a distance of a few Mpc as considered here.

\begin{figure}
    \centering
    \begin{subfigure}[b]{0.49\textwidth}
        \centering
        \includegraphics[width=\textwidth]{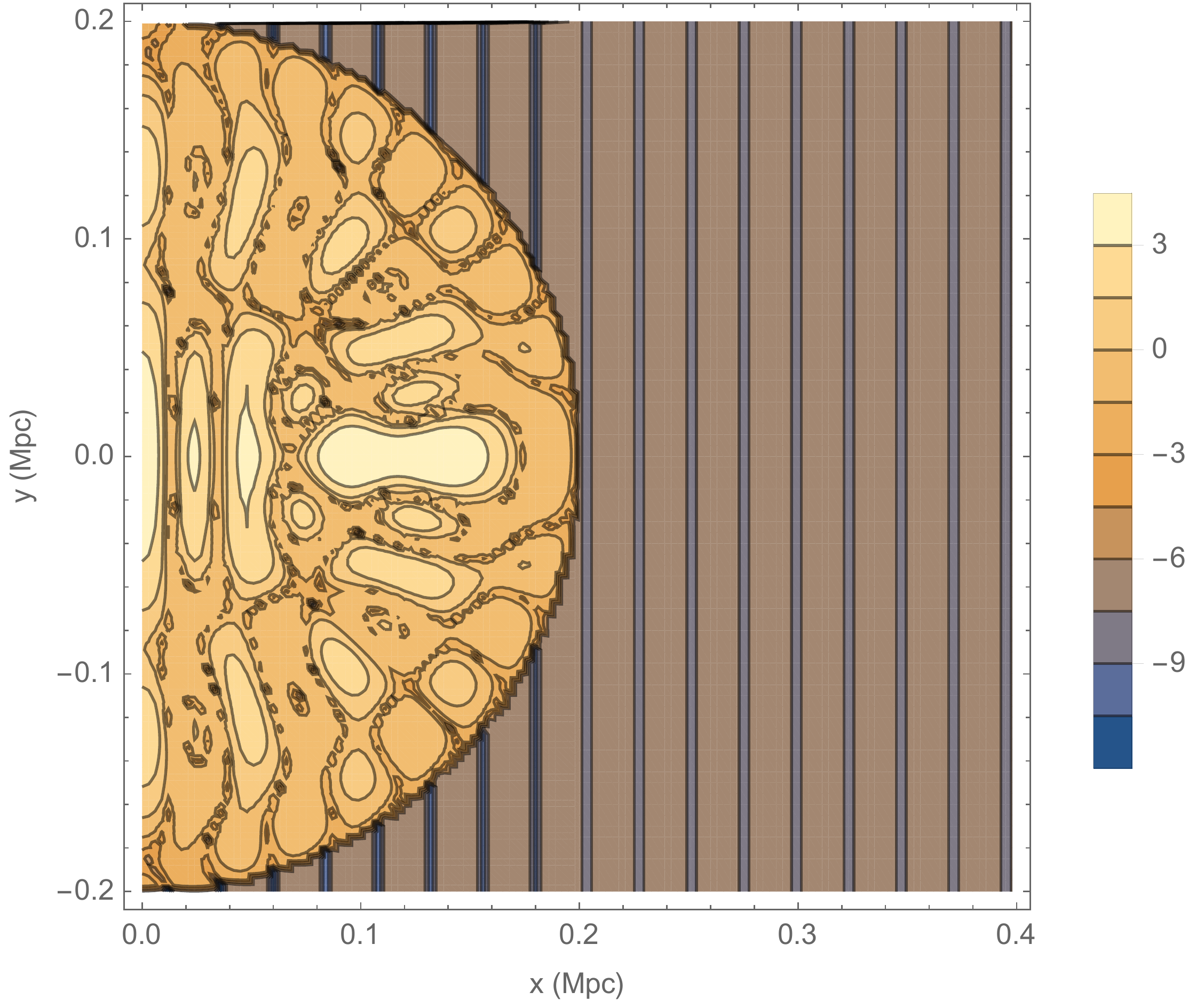}
        \caption{$\ln\DGO\big|_\tn{SS}$}\label{fig:symfilledcontour}
    \end{subfigure}
\hfill
    \begin{subfigure}[b]{0.49\textwidth}
        \centering
        \includegraphics[width=\textwidth]{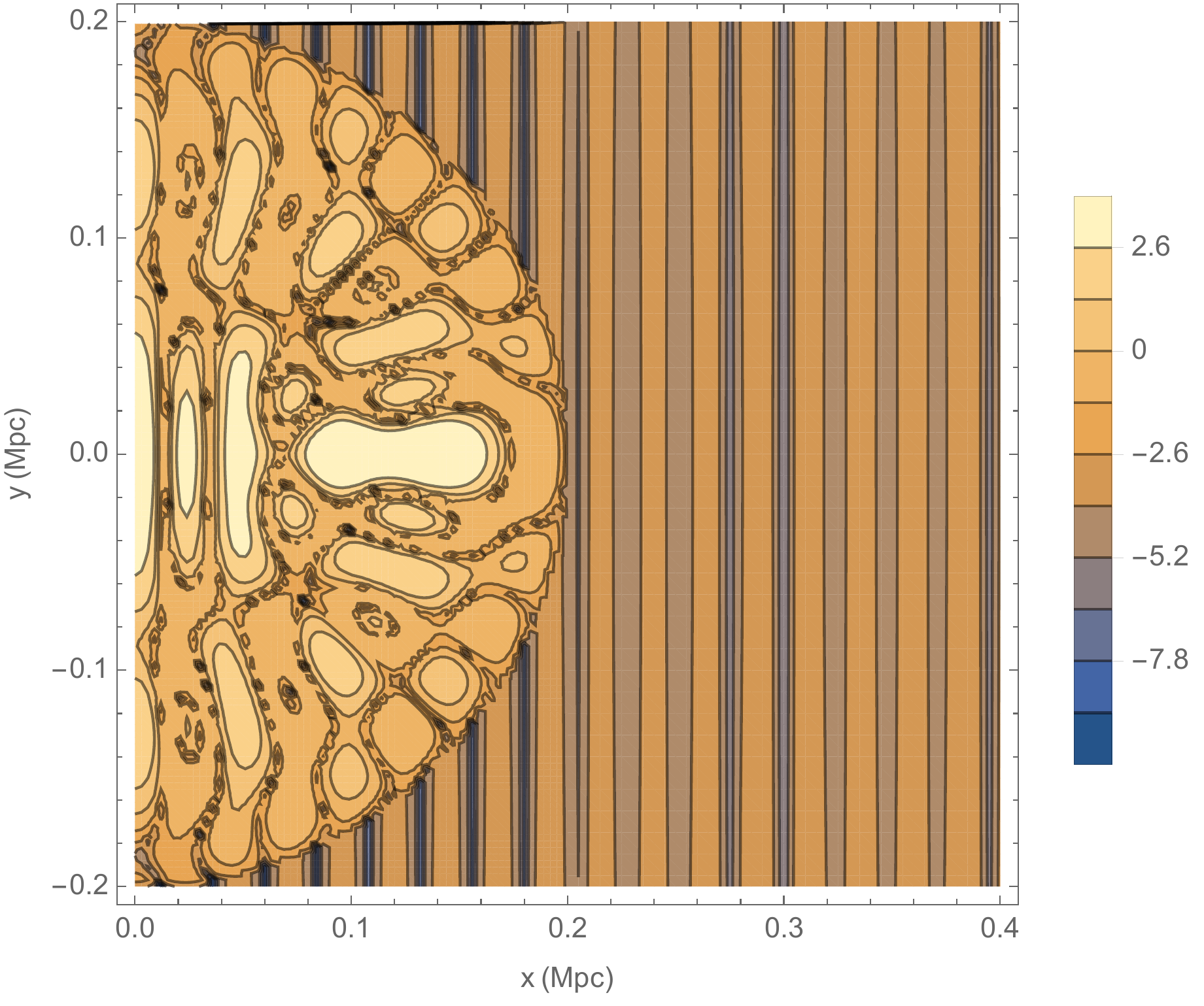}
          \caption{$\ln\DGO\big|_\tn{halo}$} \label{fig:motacontour}
    \end{subfigure}
    \caption{The fractional modulation $\DGO$ of the PPN $\gamma$ parameter by an incoming plane wave in the symmetron model, for the two different definitions of the PPN parameter \refeq{gammaSS} (left) and \refeq{gammahalo} (right), respectively. The center of the halo resides at $x=y=0$, while the Solar System is located at $r = \sqrt{x^2+y^2} = r_\tn{SS} = 0.008$~Mpc. Here, $t$ is set to $0$.}\label{fig:contourplotssymmetron}
\end{figure}

Figure \ref{fig:symfilledcontour} shows the natural logarithm of the fractional correction to the PPN $\gamma$ parameter, i.e.
\begin{equation}
  \ln \DGO = \ln\left|\frac{|\gamma-1|_\tn{non-QS}-|\gamma-1|_\qs}{|\gamma-1|_\qs}\right|\,,
  \label{eq:loggo}
\end{equation}
for planar incoming wave in the symmetron model. Here $t$ is set to $0$, and we take the real part of the solution, as written in \refeq{symplanelin}. The wave fronts inside the halo are convex with respect to the halo center, echoing the form predicted in Figure \ref{fig:refraction}. Clearly, the field perturbation is not spherically symmetric around the halo center.

Quantitatively, when inserting the values used in \cite{mota}, a fractional wave incoming wave amplitude $\delta\phi/\phi_\qs = 0.01$ and $r_\tn{BC} = 6$~Mpc (although this value is irrelevant for the plane wave), we obtain $\frac{\delta \phi}{\phi_\qs} \big |_{r_\text{SS}}\sim 4$ at the Solar System radius. This implies that the deviation of the PPN parameter $\go$ is amplified by a factor of $\left(\frac{\delta \phi}{\phi_\qs} \right)^2 \sim 16$ times over the quasi-static value $\go_\tn{QS}(r_\tn{SS})$. This amplification holds regardless of which definition of the PPN parameter is used [\refeq{gammahalo} or \refeq{gammaSS}]. For reference, using \refeq{gammahalo}, we approximately have $\go_\tn{QS}(r_\tn{SS})  \sim 10^{-6}$ for the tophat halo. 
Note that the mass distribution within the halo affects the gravitational potential at $r_\tn{SS}$ via the associated gravitational potential at that point, and hence affects the PPN parameter.  For an NFW profile \cite{NavFreWhi97}, we obtain $\go_\tn{QS}(r_\tn{SS}) \sim  7\times 10^{-8}$, in good agreement with \cite{mota}, which is smaller than in the tophat case due to the higher density in the inner regions. 
Clearly, the enhancement of $\go$ should not be taken as a literal estimate, since our linear perturbative treatment formally breaks down when $\delta\phi/\phi_\qs$ becomes of order unity. Moreover, a tophat density profile is clearly not realistic. A fractional wave amplitude of 0.01 is near the upper end of the range expected from waves generated during structure formation in symmetron models.  Llinares and Mota \cite{Llinares:2013jua} found that for their particular given model parameters and at some locations within the simulation box, cosmologically generated scalar waves can be such that these amplitudes are reached at today's epoch ($a=1$). 
All these caveats notwithstanding, we see that the wave amplitude adopted in \cite{mota} does not lead to a violation of the Cassini bound when the field configuration is a plane wave.\\

It is instructive to compare these results to the case of a spherical incoming wave, in which case the solution reads
\begin{equation}
\delta \phi  =   \mathcal{A} \, r_\tn{BC}  \sin wt \frac{\,\sin (k\, R_h) \sin (\sqrt{k^2 -m_\tn{eff,in}^2}\, r) }{\sin (k\, r_\tn{BC}) \sin (\sqrt{k^2 -m_\tn{eff,in}^2}\, R_h)}\cdot  \frac{1}{r}.
\end{equation}
This is the setup studied numerically in \cite{mota}. 
Note the dependence of the spherical solution on the combination $\mathcal{A} r_\tn{BC}$, which follows from energy conservation. A coherent spherical incoming wave leads to a strong focusing effect around the origin (in this case, the center of the halo). 
This is in contrast with the planar case, where the plane wave (with negligible $m_\tn{eff, out}$) propagates freely outside the halo.

In our linear analysis, assuming a tophat halo and at a time $t$ that maximizes the disruption (i.e. $ \sin wt=1$), we have $\frac{\delta \phi}{\phi_\qs} \sim 5\times 10^{3} $, that is, an enhancement close to three orders of magnitude larger than in the plane-wave case. Correspondingly, we formally obtain a strong violation of the Cassini bound on $\go(r_\tn{SS})$. 
The physical reason for the discrepancy is the focusing effect on the spherical wave's amplitude induced by energy conservation as $r \to 0$. This is again explicitly illustrated in the analytical solution for the spherical wave by its dependence on $\mathcal{A}\, r_\tn{BC}$.

The spherical incoming wave is the setup studied fully nonlinearly in \citep{mota} using a 1D simulation. The authors found for the same parameters that, while the enhancement is not as large as formally obtained using the linearised solution, $|\gamma(r_\tn{SS})-1|$ can take values $> 10^{-5}$, breaching the Cassini bound. Again, for the same incoming wave parameters, and changing only the wave configuration, we find that at $r_\tn{SS}$, the perturbation in the planar case is smaller by a factor $\sim 10^{-3}$ compared to the spherical case. This is a geometrical consequence and applies also to cases where the linear treatment breaks down. We thus expect that a fully nonlinear solution of the planar incoming wave will not yield a large effect on the screening in the Solar System.

\section{The chameleon case}\label{sec:cham}

The  chameleon model is another special case of the canonical scalar-tensor theory in Eq.~(\ref{eq:STA}), where now the functions $C(\phi)$ and $V(\phi)$ are chosen to be
\begin{equation}
C(\phi) = e^\frac{\beta \phi}{\mpl}, \quad V(\phi) = \frac{M^{4+n}}{\phi^n},\,\,\tn{ where we shall focus on } n=1.
\end{equation}
Our adopted values are defined in Table \ref{tab:cham}. 
The resulting field equation is then
\begin{equation}
\Box \phi = V_{\tn{eff},\phi} := -\frac{M^5}{\phi^2}+\frac{\beta }{\mpl}\rho \,e^\frac{\beta \phi}{\mpl}.
\end{equation}

In chameleon models, there is no symmetry breaking, and the field adiabatically follows its equilibrium position in the regime of large-scale structure. Hence, one does not expect large-scale waves of cosmological origin. Nevertheless, astrophysical sources of scalar waves such as Supernovae exist in these models. We will thus only consider the plane-wave case here. If the halo is screened, the quasistatic solution far inside the screening radius satisfies $V_{\tn{eff},\phi}(\phi_\qs) = 0$ \cite{daviesnfw}, which we will assume here.
It is worth emphasizing that all considerations in this section also apply to $f(R)$ gravity \cite{carroll04a,Starobinsky:2007hu}. 

\begin{table}[b]
\caption{A complete list of the quantities, including their definitions, notations and adopted values, that appear in the chameleon model, Section \ref{sec:cham}.}\label{tab:cham}
\begin{tabular}{ p{2cm} p{9cm} p{3cm}   }
 \hline
 Symbol & Definition & Value\\
\hline  
  \hline\multicolumn{3}{|c|} {\textbf{Model parameters and quantities:}}\\\hline
$\beta$ &Dimensionless coupling constant&  $1$ \\
$M$& Mass scale with units of mass, defined via $V(\phi)$ &$2\, \tn{meV}$ \cite{beta}\\
  $m_\tn{eff,in}$ & Effective field mass inside tophat halo at $\phi_\tn{QS}$.
   & $1.5 \times 10^{-30} \,\tn{GeV}/c^2$ \\
  & & $\approx 2 \times 10^{8} \,\tn{Mpc}^{-1}$\\
$m_\tn{SS}$ & Effective field mass of the scalar field in the SS. This is to be distinguished from  $m_\tn{eff,in}$ defined above. &\\ 
$V(\phi)$&Bare potential  & $M^5\phi^{-1}$\\
$V_\tn{eff}(\phi)$&Effective potential, $V(\phi)+\rho \,e^{\beta\phi/\mpl}$  & \\
  $\phi_\tn{QS}$ & Static field value minimizing $V_\tn{eff}$ inside the halo: $ \sqrt{\frac{M^5 \mpl}{\beta\rho_h}}$. Valid for  $r\in (0,R_\tn{roll})$. \cite{daviesnfw} & $3.1\,\tn{GeV}$\\
$\phi_c$ & Field at cosmic mean density: $\sqrt{\frac{M^5 \mpl}{\beta\rho_c}}$& $44\,\tn{GeV}$\\
\hline\\
\end{tabular}
\end{table}

The linearized system then takes the same form as Eq.~(\ref{linearsymmetron}) with  a different $m_\tn{eff}^2$, such that $m_\tn{eff,out}^2\approx 0 $ outside the halo while inside,
\begin{equation}
  m_\tn{eff,in}^2 \equiv m_\tn{eff}^2(\phi_\qs)\,; \quad
  m_\tn{eff}^2(\phi) =  V_{\tn{eff},\phi\phi}(\phi)\,.
\end{equation}
Far inside the screening radius, the solution is consequently the same as \refeq{symplanelin},
\begin{equation}\label{eq:champlanelin1}
\delta\phi  
= \tn{Re}\left[\mathcal{A}   e^{-i w t}\sum_{L=0}^\infty    (2L+1) i^L \frac{ j_L(k \,R_h)}{ j_L(\sqrt{k^2 -m_\tn{eff,in}^2}\,R_h)}   \,  j_L(\sqrt{k^2-m_\tn{eff,in}^2 }\, r) P_L(\cos \theta)\right] .
\end{equation}
Here we have assumed that the incoming scalar field jumps from being massless to acquiring an effective mass of $m_\tn{eff,in}$ instantaneously at $R_h$, as in the symmetron case, neglecting the thin shell of the halo. This is sufficient given our simplified setup.
Treating the Sun as a test mass, the Solar System $\gamma$ parameter analogous to \refeq{gammaSS} is 
\begin{align}
\gamma-1\Big|_\text{SS}  =\:&
-\frac{4 \beta ^2}{2 \beta ^2+e^{m_\tn{SS} r_\tn{sys}}} \,,
\label{eq:gammaSScham}
\end{align}
where $ r_\tn{sys}$ is the scale of the experiment or observation constraining $\go$, and $m_\tn{SS} = m_\tn{eff}(\phi[r_\tn{SS}])$ is the mass of the large-scale field at the Solar System location, that is, in the absence of the Sun. We have used the expressions derived by \cite{Perivolaropoulos} for massive Brans-Dicke theories. \refeq{gammaSScham} assumes that the Sun would not be screened by itself, i.e. if it were embedded in the cosmic mean density rather than the halo. If the Sun is screened, then $\gamma-1$ is further suppressed by the thin-shell factor $\delta M/M$. However, in that case we expect a negligible effect of any incoming scalar waves on the screening, since they would have to significantly modify the thin shell of the Sun. Thus, we will proceed with the assumption that the Sun is a test mass which does not screen the field by itself, for which \refeq{gammaSScham} holds. 
In that case, chameleon screening operates by making the mass of the field large (whereas the coupling remains constant). We see that, if $m_\tn{SS} \gg 1/r_\tn{sys}$, the modification to the PPN $\gamma$ parameter is exponentially damped.

For Cassini, $r_\tn{sys}$ is the distance from Saturn to Earth on the other side of the Sun. This gives  $ r_\tn{sys}\sim 11 \tn{AU}\approx 5\times 10^{-11} \tn{Mpc}$. The mass of the field given by our quasi-static solution within the smooth halo and for the fiducial model parameters would yield $m_\tn{SS} = m_\tn{eff,in} \approx 10^8 \tn{ Mpc}^{-1}$ (Table \ref{tab:cham}). However, the actual local density within the Solar System is much higher than the mean halo density assumed here, and the local value $m_\tn{SS}$ is correspondingly expected to be several orders of magnitude larger \cite{daviesnfw}. 
Nevertheless, since our analytical solution is only valid for a tophat density profile, and the focus of this paper is to study the propagation of waves of screened fields in the halo, we will continue to use $m_\tn{eff}$ as fiducial value in the following.

In order to study the wave propagation within the halo, we now consider separately the two cases $k^2 \lesssim m_\tn{eff,in}^2$ and $k^2 \gtrsim m_\tn{eff,in}^2$. 
For $k^2 \lesssim m_\tn{eff,in}^2 $, one can show that the wave amplitude is strongly suppressed inside the halo, such that the Cassini bound is not spoiled for any reasonable wave amplitudes. In this regime, \refeq{champlanelin1} becomes
\begin{equation}\label{eq:champlanelin}
\delta\phi  
= \tn{Re}\left[\mathcal{A}     e^{-i w t}\sum_{L=0}^\infty    (2L+1) i^L j_L(k \,R_h)\frac{  i_L(\sqrt{m_\tn{eff,in}^2 -k^2}\, r) }{ i_L(\sqrt{m_\tn{eff,in}^2 -k^2}\,R_h)}   \,P_L(\cos \theta)\right] \,,
\end{equation}
where $i_L$ is the modified spherical Bessel function of the first kind of order $L$, which is related to $j_L$ by $i_L(x) = i^{-l} j_L(ix) $. 
It has been shown in previous literature \cite{landau} that $|j_L(x)|<0.64 \,x^{-5/6}, \,\,\forall L,\,x\in \mathbb{R}_{>0}$. It has also been shown in \cite{paris} that $\frac{I_L (x_{-})}{I_L(x_{+})}<\left(\frac{x_{-}}{x_{+}}\right)^L,\,\, \tn{for } 0<x_{-}<x_{+}\,\tn{ and } L>-\frac{1}{2}$, where $i_L(x):=\sqrt{\frac{\pi}{2x} } I_{L+1/2}(x)$. These relations bound \refeq{champlanelin} to be numerically strongly suppressed compared to the quasi-static solution. 
This can also be confirmed numerically. 
Thus, long-wavelength chameleon waves cannot penetrate a screened halo, and hence do not disrupt the screening. This is analogous to electromagnetic waves which cannot enter a plasma if they are below the plasma frequency.

We now consider the opposite limit, $k^2 \gtrsim m_\tn{eff,in}^2 $, which is the relevant regime for astrophysical events such as Supernovae. For example, estimating $k \sim \pi/R_\star$, where $R_\star \approx$ $2.3\times 10^7 \tn{ km}$\cite{sn1987a}, is the stellar radius before collapse, we have $k\sim 10^{12}\tn{ Mpc}^{-1}\gg m_\tn{eff,in}$. 
Unfortunately, since $m_\tn{eff,in} R_h \gg 1$ numerically, high-order spherical Bessel functions contribute significantly. This makes the numerical evaluation of the solution Eq.~(\ref{eq:champlanelin1}) very difficult. 
However, when $k^2 \gg m_\tn{eff,in}^2$, as for the supernova referenced here, we can neglect $m_\tn{eff,in}$ entirely and just consider a massless field perturbation on top of the static background. This is because $k$ and $m_\tn{eff,in}$ are the only relevant scales in the wave propagation, and any corrections to the massless case are suppressed by $m_\tn{eff,in}^2/k^2$. 
In other words, a very high-frequency wave of the chameleon field is unscreened inside the (smooth) halo.

Now consider the effect of such waves within the Solar System. In order for them to propagate within the Solar System, we require $k \gtrsim m_\tn{SS}$. For a chameleon model that passes Solar System constraints in the static case however, we have $m_\tn{SS} \gg 1/r_\tn{sys}$, which implies $k \gg 1/r_\tn{sys}$ in order for a wave to propagate. Such high-frequency waves have periods that are much less than the time it takes light to travel the distance $r_\tn{sys}$. This suppresses observable effects within the Solar System (e.g. from the Cassini probe), as these average over many wave cycles leading to a cancellation. Thus, with the possible exception of a narrow window for models in which the Solar System is just barely screened, propagating scalar waves in chameleon models cannot lead to observable effects on Solar System scales.

\section{Conclusions}
\label{sec:conc}

Allowing scalar waves to propagate in scalar-tensor theories by relaxing the quasi-static assumption could have interesting consequences for the screening of the Solar System by the Milky Way halo. This could result in a tightening of the parameter space for various modified gravity models. Here we have considered a symmetron model with parameters adopted in \cite{mota}. We have studied the impact of both planar and spherical incoming waves, where the former is the more realistic case, although the latter case could arise in certain special locations due to waves generated by rapid switching of vacua of the large-scale field. We found that planar incoming waves are significantly less disruptive than their spherical counterparts considered in \cite{mota}. This is of purely geometrical origin, due to the focusing inherent in spherical incoming waves, and does not rely on the linear analysis performed here. We have further found that the field configuration generated inside the halo by an incoming plane wave is certainly not spherically symmetric about the halo center. 

The analytical approach, albeit linear, allows for a clearer understanding of the mechanisms at work. It is worth noting that for spherical waves as considered in \cite{mota}, the perturbation is explicitly dependent on the combination $A \, r_\tn{BC}$, a constant, due to energy conservation as the spherical wave closes in. This means that the incoming wave's amplitude is greatly magnified by the time it reaches the halo. Note that $r_\tn{BC}$ corresponds to $r_\tn{max}$, which is the edge of the simulation in the study of \cite{mota} and as such corresponds to an arbitrary choice. Nevertheless, waves do propagate inside the halo in symmetron models, and could potentially have consequences for models that are just marginally screened.

We also studied waves in chameleon models, which have been claimed to also potentially suffer from disruptions of screening by incoming scalar waves. In this case, the scalar waves are necessarily of astrophysical origin and therefore planar in nature. For $k\lesssim m_\tn{eff}$, we show analytically that the effect of waves on the screening in the Solar System is strongly suppressed, while it is numerically difficult to compute $\delta\phi$ for $k>m_\tn{eff}$. Since the only relevant scales are $k$ and the mass of the field $m_\tn{eff}$ inside the halo, it is clear however that if $k\gg m_\tn{eff}$, the incoming scalar wave would effectively be propagating freely within the halo. For any viable chameleon model in which the Solar System is screened, these waves would thus have to have periods that are much smaller than the light travel time through the Solar System. This means that their effects on observables such as time delay are strongly suppressed. Thus, we can effectively rule out significant observable effects of chameleon waves on Solar System scales. The crucial difference from the symmetron case is that the chameleon screening operates by making the field massive, thus limiting the propagation of low-frequency waves, while the symmetron screening operates by suppressing the coupling to matter.

To summarize, while disruptions of the screening of the Solar System by scalar waves are in principle possible in symmetron models, the effects are much less severe than previously thought when realistic wave configurations are considered. Going forward, it would be interesting to look into the influence of the halo's density distribution on the wave propagation. The dependence of the effect on the screening upon model and incoming wave parameters should also be investigated further.
On the other hand, we were able to rule out any such effects for chameleon models.

\section{Acknowledgment}
SI thanks Robert Hagala, Claudio Llinares, Razieh Pourhasan, Amol Upadhye, Alessandra Silvestri, Alexandre Barreira, Simon Nakach and Ian Jubb for helpful discussions.

FS acknowledges support from the Marie Curie Career Integration Grant  (FP7-PEOPLE-2013-CIG) ``FundPhysicsAndLSS,'' and Starting Grant (ERC-2015-STG 678652) ``GrInflaGal'' from the European Research Council.

\bibliographystyle{jhep}
\bibliography{ref}

\providecommand{\href}[2]{#2}\begingroup\raggedright\begin{thebibliography}{10}

\bibitem{cassini}
B.~Bertotti, L.~Iess and P.~Tortora, \emph{A test of general relativity using
  radio links with the cassini spacecraft},
  \href{http://dx.doi.org/10.1038/nature01997}{\emph{Nature} {\bf 425} (sep,
  2003) 374--376}.

\bibitem{Khoury03}
J.~Khoury and A.~Weltman, \emph{{Chameleon cosmology}},
  \href{http://dx.doi.org/10.1103/PhysRevD.69.044026}{\emph{Phys. Rev.} {\bf
  D69} (2004) 044026}, [\href{http://arxiv.org/abs/astro-ph/0309411}{{\tt
  astro-ph/0309411}}].

\bibitem{khourysym}
K.~Hinterbichler and J.~Khoury, \emph{{Symmetron Fields: Screening Long-Range
  Forces Through Local Symmetry Restoration}},
  \href{http://dx.doi.org/10.1103/PhysRevLett.104.231301}{\emph{Phys. Rev.
  Lett.} {\bf 104} (2010) 231301}, [\href{http://arxiv.org/abs/1001.4525}{{\tt
  1001.4525}}].

\bibitem{Llinares:2013jua}
C.~Llinares and D.~F. Mota, \emph{{Cosmological simulations of screened
  modified gravity out of the static approximation: effects on matter
  distribution}},
  \href{http://dx.doi.org/10.1103/PhysRevD.89.084023}{\emph{Phys. Rev.} {\bf
  D89} (2014) 084023}, [\href{http://arxiv.org/abs/1312.6016}{{\tt
  1312.6016}}].

\bibitem{mota}
R.~Hagala, C.~Llinares and D.~F. Mota, \emph{Cosmic tsunamis in modified
  gravity: Disruption of screening mechanisms from scalar waves},
  \href{http://dx.doi.org/10.1103/PhysRevLett.118.101301}{\emph{Phys. Rev.
  Lett.} {\bf 118} (Mar, 2017) 101301}.

\bibitem{h}
{\scshape Planck} collaboration, P.~A.~R. Ade et~al., \emph{{Planck 2013
  results. XVI. Cosmological parameters}},
  \href{http://dx.doi.org/10.1051/0004-6361/201321591}{\emph{Astron.
  Astrophys.} {\bf 571} (2014) A16},
  [\href{http://arxiv.org/abs/1303.5076}{{\tt 1303.5076}}].

\bibitem{will}
C.~M. Will, \emph{The confrontation between general relativity and experiment},
  \href{http://dx.doi.org/10.12942/lrr-2014-4}{\emph{Living Reviews in
  Relativity} {\bf 17} (jun, 2014) }.

\bibitem{assb}
R.~Hagala, C.~Llinares and D.~F. Mota, \emph{Cosmological simulations with
  disformally coupled symmetron fields},
  \href{http://dx.doi.org/10.1051/0004-6361/201526439}{\emph{Astronomy {\&}
  Astrophysics} {\bf 585} (dec, 2015) A37}.

\bibitem{1512.00615}
J.~Ã. Lindroos, C.~Llinares and D.~F. Mota, \emph{{Wave Propagation in Modified
  Gravity}}, \href{http://dx.doi.org/10.1103/PhysRevD.93.044050}{\emph{Phys.
  Rev.} {\bf D93} (2016) 044050}, [\href{http://arxiv.org/abs/1512.00615}{{\tt
  1512.00615}}].

\bibitem{qssym}
K.~Hinterbichler, J.~Khoury, A.~Levy and A.~Matas, \emph{Symmetron cosmology},
  \href{http://dx.doi.org/10.1103/physrevd.84.103521}{\emph{Physical Review D}
  {\bf 84} (nov, 2011) }.

\bibitem{NavFreWhi97}
J.~F. Navarro, C.~S. Frenk and S.~D.~M. White, \emph{{A Universal Density
  Profile from Hierarchical Clustering}},
  \href{http://dx.doi.org/10.1086/304888}{\emph{Astrophys. J.} {\bf 490} (1997)
  493--508}, [\href{http://arxiv.org/abs/astro-ph/9611107}{{\tt
  astro-ph/9611107}}].

\bibitem{daviesnfw}
R.~Pourhasan, N.~Afshordi, R.~B. Mann and A.~C. Davis, \emph{{Chameleon
  Gravity, Electrostatics, and Kinematics in the Outer Galaxy}},
  \href{http://dx.doi.org/10.1088/1475-7516/2011/12/005}{\emph{JCAP} {\bf 1112}
  (2011) 005}, [\href{http://arxiv.org/abs/1109.0538}{{\tt 1109.0538}}].

\bibitem{carroll04a}
S.~M. {Carroll}, V.~{Duvvuri}, M.~{Trodden} and M.~S. {Turner}, \emph{{Is
  cosmic speed-up due to new gravitational physics?}},
  \href{http://dx.doi.org/10.1103/PhysRevD.70.043528}{\emph{\prd} {\bf 70}
  (Aug., 2004) 043528--+},
  [\href{http://arxiv.org/abs/arXiv:astro-ph/0306438}{{\tt
  arXiv:astro-ph/0306438}}].

\bibitem{Starobinsky:2007hu}
A.~A. Starobinsky, \emph{{Disappearing cosmological constant in f(R) gravity}},
  \href{http://dx.doi.org/10.1134/S0021364007150027}{\emph{JETP Lett.} {\bf 86}
  (2007) 157--163}, [\href{http://arxiv.org/abs/0706.2041}{{\tt 0706.2041}}].

\bibitem{beta}
C.~Burrage and J.~Sakstein, \emph{A compendium of chameleon constraints},
  \href{http://dx.doi.org/10.1088/1475-7516/2016/11/045}{\emph{Journal of
  Cosmology and Astroparticle Physics} {\bf 2016} (nov, 2016) 045--045}.

\bibitem{Perivolaropoulos}
L.~Perivolaropoulos, \emph{{PPN Parameter gamma and Solar System Constraints of
  Massive Brans-Dicke Theories}},
  \href{http://dx.doi.org/10.1103/PhysRevD.81.047501}{\emph{Phys. Rev.} {\bf
  D81} (2010) 047501}, [\href{http://arxiv.org/abs/0911.3401}{{\tt
  0911.3401}}].

\bibitem{landau}
L.~J. Landau, \emph{Bessel functions: Monotonicity and bounds},
  \href{http://dx.doi.org/10.1112/S0024610799008352}{\emph{Journal of the
  London Mathematical Society} {\bf 61} (2000) 197--215}.

\bibitem{paris}
A.~Laforgia, \emph{Bounds for modified bessel functions},
  \href{http://dx.doi.org/10.1016/0377-0427(91)90087-z}{\emph{Journal of
  Computational and Applied Mathematics} {\bf 34} (apr, 1991) 263--267}.

\bibitem{sn1987a}
F.~Taddia, M.~D. Stritzinger, J.~Sollerman, M.~M. Phillips, J.~P. Anderson,
  M.~Ergon et~al., \emph{The type {II} supernovae 2006v and 2006au: two
  {SN}~1987a-like events},
  \href{http://dx.doi.org/10.1051/0004-6361/201118091}{\emph{Astronomy {\&}
  Astrophysics} {\bf 537} (jan, 2012) A140}.

\end{thebibliography}\endgroup

\end{document}